\documentclass[apj]{emulateapj}

\usepackage[letterpaper,portrait,left=1.25in,right=0.75in,top=1.5in,bottom=0.5in]{geometry}                
\usepackage{graphicx}
\usepackage{amssymb}
\usepackage{amsmath}
\usepackage{epstopdf}
\usepackage[all]{nowidow}

\usepackage[backref,breaklinks,colorlinks,citecolor=blue]{hyperref}
\usepackage[figure,figure*,table]{hypcap}

\usepackage{natbib}
\let\oldenddeluxetable\enddeluxetable
\let\olddeluxetable\deluxetable
\makeatletter
\renewenvironment{deluxetable}[1]{
\olddeluxetable{[#1]}
\def\pt@format{#1}%
}{\oldenddeluxetable}
\makeatother

\defcitealias{2005ApJS..159..141V}{VF05}
\defcitealias{Brewer:2016uh}{B16}
\defcitealias{2015ApJ...805..126B}{B15}

\newcommand{\teff}{$T_{\mathrm{eff}}$}
\newcommand{\logg}{$\log g$}

\newcommand{\vsini}{$v \sin i$}
\newcommand{\vmac}{$v_{\rm mac}$}

\begin{document}
\title{C/O and Mg/Si Ratios of Stars in the Solar Neighborhood}

%
\author{John M. Brewer, Debra A. Fischer}
	\affil{Department of Astronomy, Yale University}
	\affil{260 Whitney Avenue, New Haven, CT 06511, USA}
	\email{john.brewer@yale.edu}
	\email{debra.fischer@yale.edu}
%


\begin{abstract}
The carbon to oxygen ratio in a protoplanetary disk can have a dramatic influence on the compositions of any terrestrial planets formed.  In regions of high C/O, planets form primarily from carbonates and in regions of low C/O, the ratio of magnesium to silicon determines the types of silicates which dominate the compositions.  We present C/O and Mg/Si ratios for 849 F, G, and K dwarfs in the solar neighborhood.  We find that the frequency of carbon-rich dwarfs in the solar neighborhood is $< 0.13\%$ and that 156 known planet hosts in the sample follow a similar distribution as all of the stars as a whole.  The cosmic distribution of Mg/Si for these same stars is broader than the C/O distribution and peaks near 1.0 with $\sim 60\%$ of systems having $1 \leq$~Mg/Si~$< 2$, leading to rocky planet compositions similar to the Earth.  This leaves $40\%$ of systems that can have planets that are silicate rich and may have very different compositions than our own.
\end{abstract}


\maketitle

%
\section{Introduction} \label{sec;introduction}
Obtaining the masses of transiting exoplanets allows us to measure their bulk densities, an important step in understanding their interior compositions.  Improvements in radial velocity precision are pushing those measurements into the regime of terrestrial mass planets \citep{Motalebi:2015bf,Pepe:2011jc}.  A range of possible interiors is still possible due to  varying ratios of heavy elements \citep{2012ApJ...759L..40M}.

Chemistry is one of the important environmental factors in the protoplanetary disk.  Given a temperature and pressure profile for the disk, the abundance of different elements determines the composition and structure of planet interiors.  Because stellar atmospheres evolve slowly, the elemental abundances of planet host stars reflects the composition of their planet forming disks.

The prevalence of carbon and oxygen and the favorable thermodynamics of carbon monoxide bonds make CO a dominant molecular component of the gaseous protoplanetary disk \citep{2014prpl.conf..363P,2014pccd.book.....G}. As solids condense out, the ratio of carbon to oxygen influences the planetesimal geology \citep{2014prpl.conf..363P}.  If the C/O ratio is above a threshold value, then there is almost no free oxygen available to form silicates and the geology will be dominated by carbonates \citep{2005astro.ph..4214K}; under high pressure, these are expected to form crystal diamond compositions \citep{2012ApJ...759L..40M}.  When the C/O ratio is below that threshold, planetesimal geology is primarily magnesium silicates.  The threshold ratio is $\sim 1$ at the formation sites of the planetesimals, but due to the temperature structure and evolution of the protoplanetary disk, initial ratios as low as 0.65 can still lead to carbon rich conditions \citep{Moriarty:2014cb}. The solar system C/O is $\sim 0.54$ \citep{2011SoPh..268..255C,2009ARA&A..47..481A} and the rocky planets have magnesium silicate compositions despite carbon being almost ten times more abundant in the solar photosphere than silicon \citep{2009ARA&A..47..481A}.

The nucleosynthetic pathways for carbon and oxygen are slightly different.  Although both are formed in the explosions of massive stars, the majority of carbon comes instead from lower mass stars \citep{Woosley:2002ck}. The evolutionary pathway to carbon rich dwarf is exceedingly narrow and involves mass transfer from an AGB companion \citep{2005astro.ph..2152S}.  Oxygen is the most abundant heavy element produced by massive stars \citep{1995ApJS..101..181W}.  Their more rapid evolution will result in an initially low C/O ratio that will increase as AGB winds contribute carbon to the interstellar medium.  Galactic chemical evolution (GCE) models predict that the ratio of C/O should peak at just slightly greater than solar \citep{2015ApJ...804...40G,2006ApJ...653.1145K,1999A&A...342..426G}.  Photometric surveys and targeted searches for carbon dwarfs \citep{2005astro.ph..2152S} have found that fewer than 0.1\% of main sequence stars are carbon rich (i.e. C/O~$> 1$) \citep{2008AJ....136.1778C}.

In contrast, some high resolution spectral analyses of carbon and oxygen lines have found that the fraction of stars with C/O ratios $> 1$ may be as high as 6-10\% \citep{2011ApJ...735...41P,2010ApJ...725.2349D} of all stars. Spectroscopic measurements of both elements are difficult due to blends and the small number of atomic lines of C and O in the visible. More recent analyses with updated line parameters have found lower values \citep{2014A&A...568A..25N,2014ApJ...788...39T}.  Uncertainties in the high C/O tail of the distribution could also reduce the number of stars with high C/O to 1-5\% \citep{2012ApJ...747L..27F}. Almost constant errors in the logarithmic abundances lead to a log normal distribution of the C/O errors. The result is higher uncertainties at higher C/O ratios.  However, 1-5\% of stars with C/O~$> 1$ is still more than a factor of 10 greater than that expected from the frequency of carbon dwarfs or from galactic chemical evolution models.  Fewer than 1\% of M dwarfs in the solar neighborhood, which should show clear evidence of high C/O ratios in their spectra, have $0.8 <$~C/O~$< 1.0$.

For oxygen rich systems (i.e. C/O~$< 1$) the silicate minerology will be determined by the local Mg/Si ratio.  Magnesium and silicon form in nearly equal abundance to each other, solar Mg/Si~$= 1.05$ \citep{Grevesse:2007cx}, and there are 3 regimes of mineral formation as that ratio varies.  When Mg/Si~$< 1$, the magnesium is consumed in producing pyroxene (MgSiO$_3$) with the remainder of the silicon in feldspars.  In systems with $1 <$~Mg/Si~$< 2$ there will be a mix of pyroxene and olivine (Mg$_2$SiO$_4$) similar to the solar system.  Olivine and other magnesium compounds such as MgO and MgS will dominate in the case of Mg/Si~$> 2$ \citep{2014pccd.book.....G,CarterBond:2012dma}. The particular geology could have profound consequences for differentiation and energy transport in the planet interior, and ultimately for the habitability of the resulting planets and our ability to characterize them.

We examine the C/O ratios for 849 F, G, and K dwarfs in the solar neighborhood. We also derive the distribution of Mg/Si ratios and compare both the C/O and Mg/Si ratios for known planet hosts to the general sample.  These distributions can inform theoretical expectations of protoplanetary disk compositions and the range of exoplanet interiors.

%
\section{Data and Analysis}
The catalog of \citet{Brewer:2016uh} has abundances of 15 elements including C, O, Mg, and Si for more than 1600 F, G, and K stars.  The stars were all observed using the HIRES instrument on the Keck telescope with the same instrumental setup.  Most of the stars were observed as part of the California Planet Search (CPS) program and have a typical SNR~$\gg 100$. A subset of fainter targets, mainly Kepler stars, were observed at lower SNR.  The analysis used Spectroscopy Made Easy (SME) in an iterative procedure with a single line list.  The initial guesses for the stellar parameters assumed a solar abundance pattern and chose \teff\ from broadband colors. Global stellar parameters  (\teff, \logg, [M/H], and \vmac\ with \vsini=0) and the abundances of three $\alpha$ elements (Ca, Si, Ti) were fit with a synthetic model using a Levenberg Marquardt algorithm.  These global parameters were then fixed when fitting for abundances of 15 elements including the three $\alpha$ elements.  Using the new abundance pattern, an iteration was carried out to refine the spectral synthesis model.  Small residual trends in abundances with \teff\ were fit and removed to provide corrected abundances.

\subsection{Solar Relative Abundances} \label{sec:solar_values}
The spectral model included more than 7500 atomic and molecular lines in segments between 5160~\AA and 7800~\AA. The line parameters for the analysis were tuned against the solar flux atlas of \citet{2011ApJS..195....6W} using the abundance pattern of \citet{Grevesse:2007cx}.  Twenty spectra of reflected sunlight from 4 asteroids taken on 5 separate epochs were used to verify that the procedure returned the solar values.  The mean offset in abundance was $\sim 0.02$~dex with an rms scatter of less than 0.01~dex for most elements.  The mean offsets in asteroid abundances were removed from the final abundances to yield self-consistent solar relative values.

\subsection{Sample Selection} \label{sec:sample_selection}
The final corrected abundances show a precipitous rise in scatter for stars with SNR~$< 100$ and a weaker, but noticeable, increase in scatter for stars with \vsini~$\gtrsim 15$~km/s and \teff~$\lesssim 5000$~K.  There are also trends of opposite sign in the oxygen and carbon abundances above $\sim$~6000~K. We therefore restrict our analysis to those with SNR~$> 100$, 6100~K~$>$~\teff~$> 4800$, and \vsini~$\leq 20$~km/s. Neither magnesium or silicon show significant trends for all stars in the \citet{Brewer:2016uh} catalog and these elements have a lower scatter than oxygen or carbon.

Our goal in this work is to determine the distribution of C/O ratios in dwarf stars in the local neighborhood in order to assess potential influences on planet formation.  In order to avoid carbon enhancement from stellar evolution, we restrict the sampe to main sequence stars with \logg~$\geq 3.5$.  These cuts resulted in a sample of 849 stars.

There are 156 known planet hosts out of the 849 stars in our sample (dark blue regions in Figure \ref{fig:co_ratio_dwarfs}).  The majority of the planet hosts were detected with radial velocity and so are biased towards more massive planets; 110 of the hosts have giant planets ($> 0.7 R_{Jup}$ or $> 0.2 M_{Jup}$) and 60 of those have semi-major axes less than 1~AU.  The remainder of the stars on the CPS Doppler survey with undetected planets are likely to host smaller planets; very few will have gas giant planets.

\subsection{Analysis}

We used the values for [C/H], [O/H], (C/H)$_{\odot}$, and (O/H)$_{\odot}$ published in \citet{Brewer:2016uh} to calculate the C/O ratios (Equation \ref{eqn:co_ratio}).  Here [X/H] is the standard notation for the log$_{10}$ of the solar relative number abundance of an element with respect to hydrogen and (X/H)$_{\odot}$ is the log$_{10}$ number abundance of the element with respect to hydrogen +12 for the Sun.

\begin{equation}  \label{eqn:co_ratio}
	X_1/X_2 = 10^{([X_1/H] + (X_1/H)_{\odot}) - ([X_2/H] + (X_2/H)_{\odot})}
\end{equation}

We derived Mg/Si ratios for the sample as well using the same procedure.  

The errors presented in the catalog are given as log relative errors for each element. This allowed a direct calculation of the relative errors from the quadrature sum of the errors for the two elements in the ratio.  The uncertainties for [C/H], [O/H], [Mg/H], and [Si/H] are 0.026 dex, 0.036 dex, 0.012 dex, and 0.008 dex respectively.  The errors were calculated using all of the stars in the sample with repeated observations.  The pairwise abundance differences for all such stars in the sample were collected and the central 68\% of the distribution for each element was taken.  These values were multiplied by a factor of 2 for the final uncertainties as discussed in detail in \citet{Brewer:2016uh}.  The reported uncertainties in [Mg/H] and [Si/H] are quite small and may not fully capture uncertainties due to NLTE effects, errors in the atmosphere models or the empirically corrected trend.  Because atmospheric models are tuned to the Sun, they are less robust for stars that are increasingly non-solar. To minimize systematic errors from model atmospheres, we established a subset of comparison stars that were similar in temperature and surface gravity to the Sun.

%
\section{Results} \label{sec:results}
\begin{figure*}[htpb!] 
   \centering
   \includegraphics[width=0.95\textwidth]{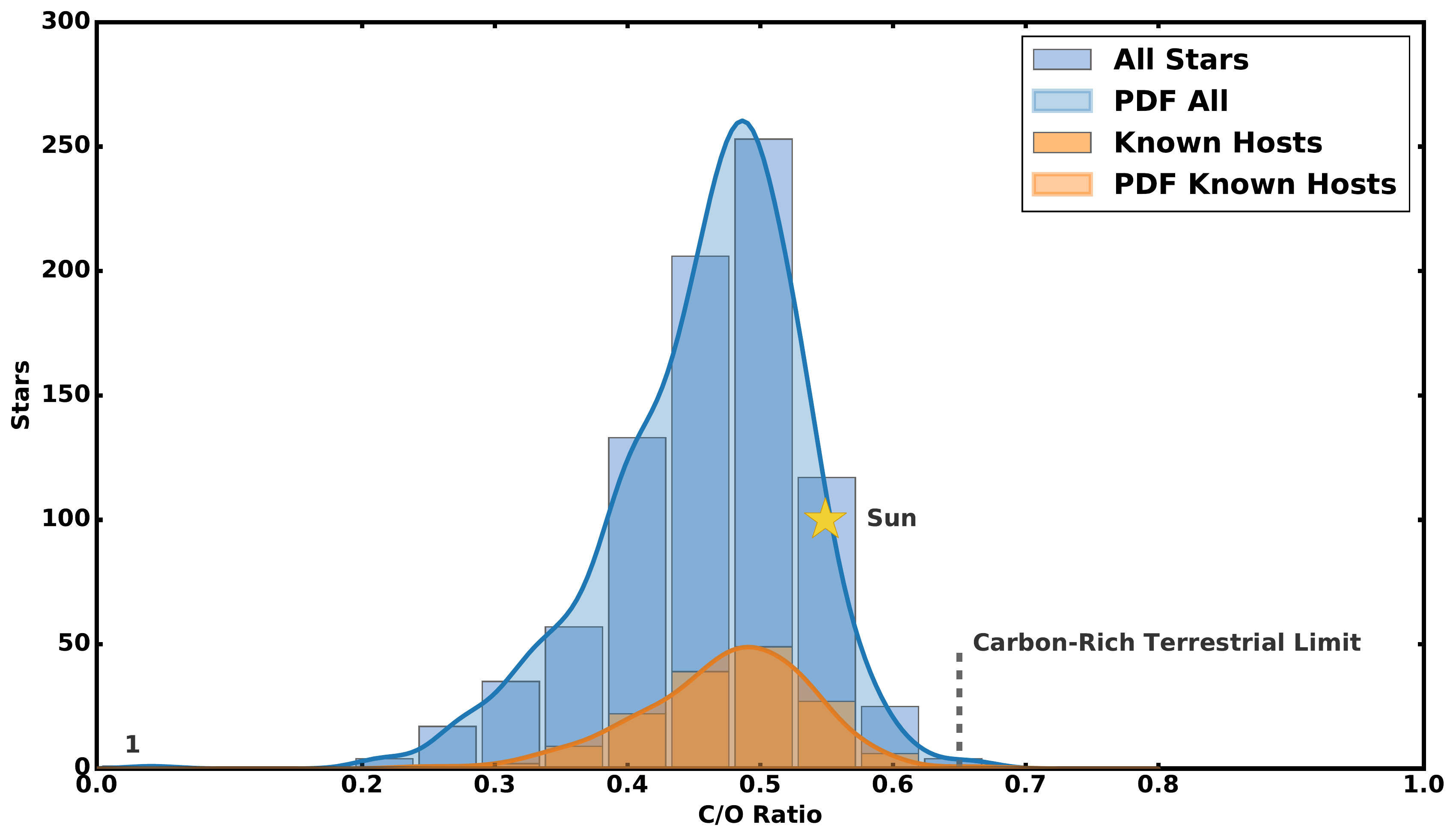} 
   \caption{The distribution of C/O ratios for a sample of 849 high SNR, slowly rotating dwarf stars from the catalog of \citet{Brewer:2016uh} with 6100~K~$>$~\teff~$> 4800$.  The orange portions of the bars indicate known planet hosts.  Probability density functions (PDF) calculated using a Gaussian kernel density estimator and scaled to the area of the histograms are also presented. The gold star marks the position of the Sun in this sample (C/O = 0.55) and the dashed line marks the C/O ratio at which it may be possible to form carbon-rich terrestrial planets. \\ }
   \label{fig:co_ratio_dwarfs}
\end{figure*}

The distribution of C/O in slowly rotating F, G, and K dwarfs in the solar neighborhood (Figure \ref{fig:co_ratio_dwarfs}) has the expected distribution from stellar models \citep{1995ApJS..101..181W} and those of galactic chemical evolution \citep{2015ApJ...804...40G}.  The maximum ratio in our sample of 849 stars is C/O~$= 0.66 \pm 0.068$. Only four stars approach the minimum C/O ratios necessary to form carbon rich rocky planets \citep{Moriarty:2014cb} and none are in the carbon rich regime that would make such planets probable. The median of the sample is 0.47, which means that the Sun at C/O = 0.55 is slightly carbon rich compared to the stars in our sample. The standard deviation in the sample is 0.07, though the distribution has a long tail towards lower C/O ratios.

\begin{figure*}[htpb!] 
   \centering
   \includegraphics[width=0.95\textwidth]{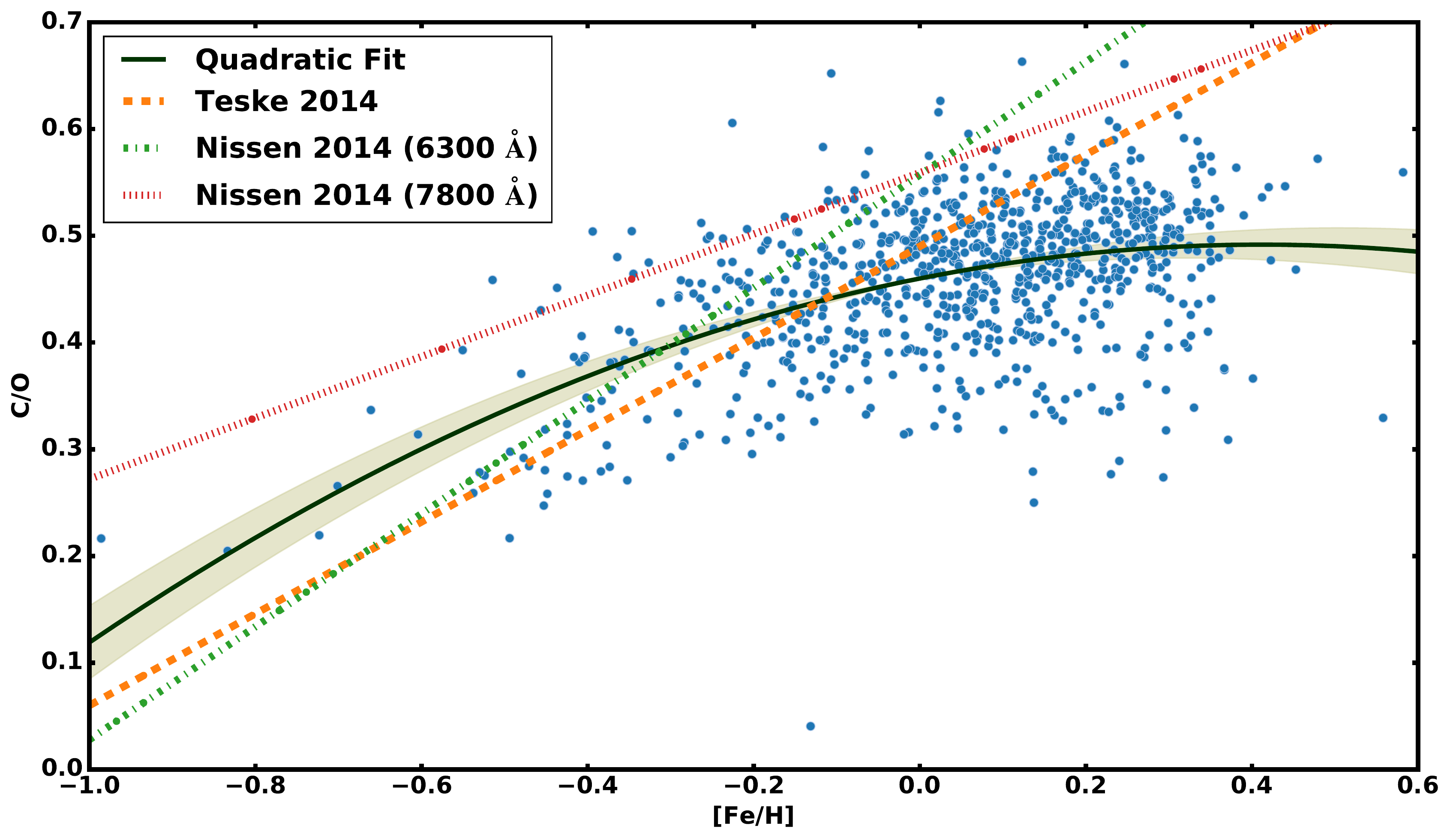} 
   \caption{The distribution of C/O as a function of [Fe/H] for all 849 stars in our sample (blue points).  The black solid line indicates a quadratic fit to the distribution and the grey region indicates the 1-$\sigma$ confidence region.  The fit shows an increase in C/O with increasing metallicity that levels off at [Fe/H]~$\gtrsim -0.1$~dex.  For comparison, we have included the linear fits of \citet{2014A&A...568A..25N} (using the 6300~\AA\ oxygen line and the 7800~\AA\ triplet) and \citet{2014ApJ...788...39T} that have similar slopes to the low metallicity end of our sample, but don't adequately capture the leveling off of C/O at higher metallicities. \\ }
   \label{fig:co_fe_compare}
\end{figure*}

At lower metallicities, the C/O ratio increases with [Fe/H], though at [Fe/H]~$\gtrsim -0.1$~dex the distribution is relatively flat.  This plateau in C/O matches expectations from the galactic chemical evolution model for stars in the solar neighborhood of \citet{2015ApJ...804...40G}, though the exact location of this plateau varies by both location in the disk and carbon and oxygen yields from low and intermediate mass stars \citep{Carigi:2005bw}.  We used Gaussian kernel regression to find an appropriate functional fit to the mean of the distribution.  The final quadratic fit (Figure \ref{fig:co_fe_compare}) closely matches the Gaussian kernel regression with a rms scatter of 0.06.  Recent results from \citet{2013A&A...552A..73N,2014A&A...568A..25N} and \citet{2014ApJ...788...39T} fit linear trends that have similar slopes to our fit at lower metallicities; however, our distribution differs in that it levels off at metallicities~$> -0.2$ and we do not understand the reason for that difference.

The C/O ratio of the metal rich star 55~Cnc has been reported to have C/O as high as 1.12 \citep{2010ApJ...725.2349D}, motivating speculation about the possibility of diamond planets \citep{2012ApJ...759L..40M}.  Subsequent analyses yield lower values for 55~Cnc, but the lowest at $0.78 \pm 0.08$ \citep{2013ApJ...778..132T} still allowed for the possibility of forming carbon rich terrestrial planets. In contrast, we derive a C/O ratio for 55 Cnc of 0.53~$\pm 0.054$, almost identical to the Sun.

\begin{figure*}[htpb!] 
   \centering
   \includegraphics[width=0.95\textwidth]{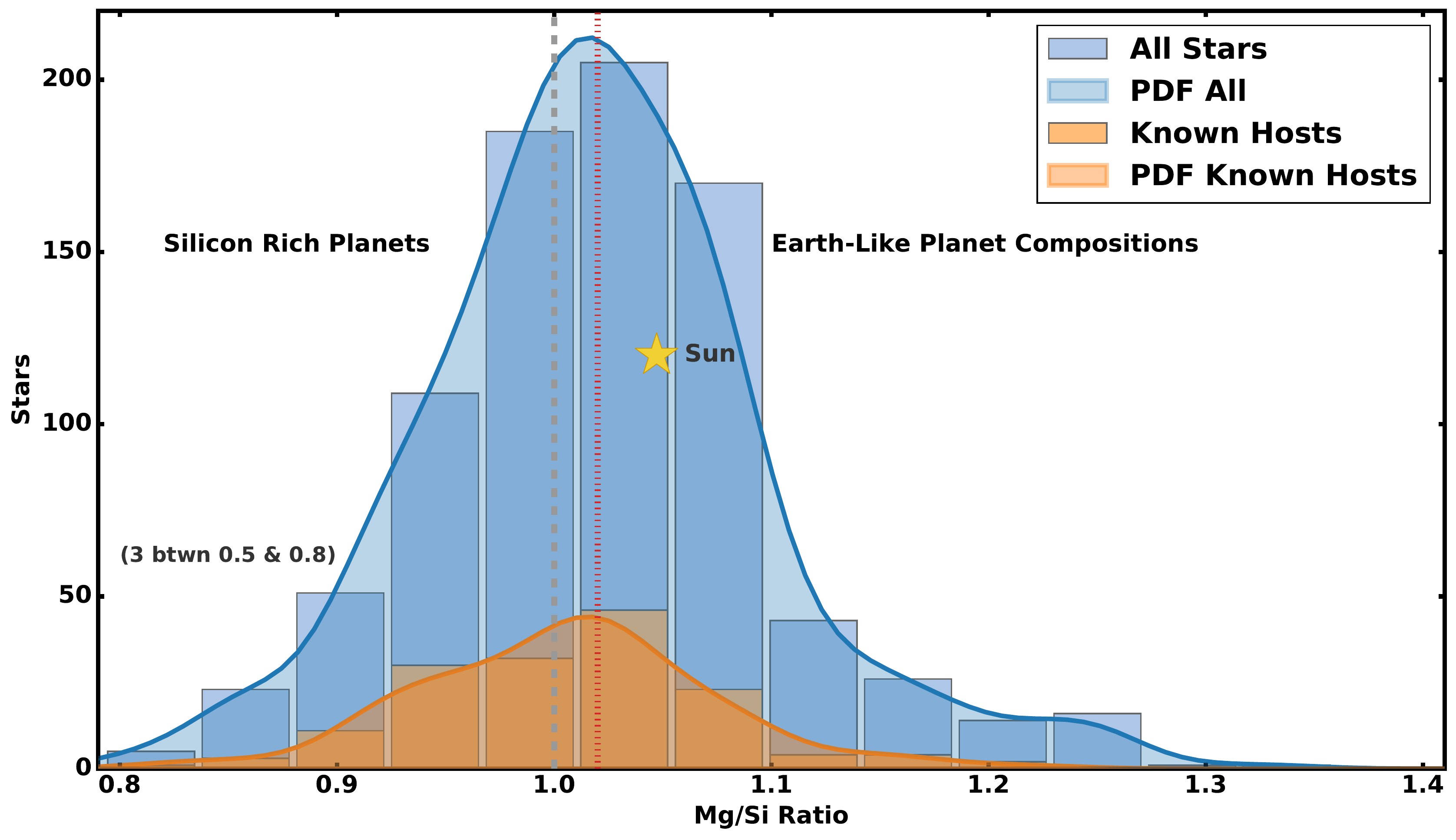} 
   \caption{The distribution of Mg/Si ratios for known planet hosts (orange) and all other stars in our sample (blue).  Probability density functions (PDF) calculated using a Gaussian kernel density estimator and scaled to the area of the histograms are also presented. The red dotted line marks the median of the sample.  The grey-dashed line marks the point at which silicate chemistry in the disk would lead to differing final compositions for rocky planets.  Below 1.0, the excess of silicon would allow additional silicon species to form beyond the more Earth-like compositions which would form in systems with Mg/Si~$\geq 1.0$.  Three stars with $0.5 < $~Mg/Si~$ < 0.8$ were excluded from the plot for clarity. \\ 
   ~ \\ }
   \label{fig:mgsi_all_stacked}
\end{figure*}

The distribution of C/O ratios for known planet hosts is very similar to that of stars without known planets.  However, we note that our knowledge about the existence of planets is strongly biased toward gas giant planets in this sample.  Since the C/O ratio is correlated with [Fe/H] (Figure \ref{fig:co_fe_compare}), we expect that a small shift in the peak of the C/O distribution for planet hosting stars statistically reflects the higher metallicity of these stars. The distribution of Mg/Si ratios has a median value of 1.02 (Figure \ref{fig:mgsi_all_stacked});  40\% having Mg/Si~$< 1$ and the remaining 60\% of the stars have Mg/Si between 1 and 1.4.

Table \ref{table:co_ratios} contains the abundance data from \citet{Brewer:2016uh} (Columns (4)-(7)) along with the calculated C/O (Column (8)), Mg/Si (Column (9)) and the number of known planets around the star (Column (10)).  Columns (1)-(3) are the SPOCS ID, name and V magnitude as presented in the catalog.  Based on the uncertainties provided in \citet{Brewer:2016uh}, the uncertainties in the C/O values are $\pm 10$\% and the uncertainties in the Mg/Si values are $\pm 3.3$\%.

\begin{center} 
\begin{deluxetable*}{ l l r r r r r r r c }

\tablecolumns{10} 
\tabletypesize{\footnotesize} 
\tablecaption{Stellar C/O and Mg/Si Ratios \label{table:co_ratios}} 
\tablehead{ 
\colhead{} & \colhead{} & \colhead{V} & \colhead{[C/H]} & \colhead{[O/H]} & \colhead{[Mg/H]} & \colhead{[Si/H]} & \colhead{C/O} & \colhead{Mg/Si} & \colhead{Known} \\
\colhead{ID} & \colhead{Name} & \colhead{mag} & \colhead{} & \colhead{} & \colhead{} & \colhead{} & \colhead{} & \colhead{} & \colhead{Planets} \\
\colhead{(1)} & \colhead{(2)} & \colhead{(3)} & \colhead{(4)} & \colhead{(5)} & \colhead{(6)} & \colhead{(7)} & \colhead{(8)} & \colhead{(9)} & \colhead{(10)} 
} 
\startdata 
   2 & HD 105 &  7.51 &  0.073 &  0.126 & -0.012 &  0.050 &  0.48 &  0.91 & $\cdots$ \\
   4 & HD 166 &  6.07 &  0.014 &  0.085 &  0.001 &  0.054 &  0.46 &  0.93 & $\cdots$ \\
   6 & HD 377 &  7.59 &  0.122 &  0.226 &  0.040 &  0.128 &  0.42 &  0.86 & $\cdots$ \\
  10 & HD 691 &  7.95 &  0.103 &  0.131 &  0.107 &  0.167 &  0.50 &  0.91 & $\cdots$ \\
  12 & HD 1388 &  6.51 &  0.005 & -0.008 &  0.027 &  0.009 &  0.55 &  1.09 & $\cdots$ \\
  13 & HD 1461 &  6.47 &  0.131 &  0.098 &  0.156 &  0.161 &  0.58 &  1.04 &  1 \\
  22 & HD 2589 &  6.18 & -0.051 &  0.111 & -0.002 & -0.107 &  0.37 &  1.33 & $\cdots$ \\
  30 & HD 3795 &  6.14 & -0.509 & -0.192 & -0.454 & -0.395 &  0.26 &  0.91 & $\cdots$ \\
  34 & HD 4208 &  7.78 & -0.227 & -0.144 & -0.208 & -0.234 &  0.44 &  1.11 &  1 \\
  36 & HD 4203 &  8.70 &  0.306 &  0.319 &  0.306 &  0.316 &  0.52 &  1.02 &  2 \\
\enddata 
\tablecomments{Table \ref{table:co_ratios} is published in its entirety in the electronic edition of this article. A portion is shown here for guidance regarding its form and content.} 
\end{deluxetable*} 
\end{center} 

\ \\

%
\section{Discussion} \label{sec:discussion}
The importance of C/O to the interior structure of rocky planets has prompted recent spectroscopic work in identifying the distribution of C/O in stars.  The carbon to oxygen ratio is also a tracer of stellar and galactic chemical evolution and photometric surveys have previously explored this same question.  Photometric surveys have the advantage of both expanded wavelength coverage into the IR and large numbers covering tens of thousands of stars.  A study combining infrared photometry and low resolution spectroscopy found that only 4 of nearly 10,000 low mass stars had C/O~$> 1$ \citep{2008AJ....136.1778C}, though even those are thought to form from binary mass transfer instead of having primordially high C/O \citep{2005astro.ph..2152S}.  Optical spectroscopic surveys analyze a relatively small number of C and O atomic lines.  The optical lines can also be difficult to measure because of line blends and the literature shows estimates for the number of stars with primordial C/O~$> 1$ ranging from $\sim 0$ to as high as 10\% \citep{2014A&A...568A..25N,2011ApJ...735...41P}.

The distribution of C/O ratios calculated here are shifted to slightly lower values relative to distributions from other spectroscopic analysis \citep{2011ApJ...735...41P,2010ApJ...725.2349D}. However, our distribution of lower C/O ratios agrees with the large photometric searches for carbon-rich dwarfs \citep{2008AJ....136.1778C}. One possible explanation for the discrepant comparison with these other spectroscopic analyses, is that in addition to the atomic C and O transitions, our analysis also includes hundreds of molecular lines, which may make our analysis more robust.  We also exclude the [OI] line at 6300~\AA\ because it is blended with a nickel line and a weak CN line. Recent spectroscopic analyses that include the oxygen triplet at 7770 \AA\ \citep{2014A&A...568A..25N,2014ApJ...788...39T,2013A&A...552A..73N,2013ApJ...778..132T} derive lower C/O ratios, in closer agreement with our distribution.

In systems with high C/O ratios, we expect to find planets with exotic compositions.  The fact that C/O ratios for other stars are so similar to the Sun suggests it will be rare for rocky planets to be carbon rich \citep{2012ApJ...759L..40M}.

Nevertheless, the broad distribution of Mg/Si in our sample combined with the overall low C/O ratios indicates that there will still be a large diversity of terrestrial planet compositions.  Within measurement uncertainties, the known planet hosts in the sample have essentially the same abundance ratio distribution as the sample as a whole.  Our stars can be divided into two regimes of Mg/Si: Earth-like magnesium silicate compositions or more silicon rich mineralogy.  Terrestrial planets around any of these stars would be dominated by the same elements as rocky planets in the solar system (O, Fe, Mg, and Si) but the specific compositions of the rocks will be influenced by which side of Mg/Si$ = 1$ line they fall on \citep{Moriarty:2014cb,2010ApJ...715.1050B,2010ApJ...725.2349D}.

Silicon rich systems (Mg/Si~$< 1$) will have their available magnesium bound up in pyroxene \citep{2010ApJ...715.1050B} leaving silicon free to form other silicate species.  The Mg/Si ratio is between 1 and 1.4 in 60\% of the systems that we considered, suggesting that Earth-like compositions may be relatively common. However, rocky planets around stars with Mg/Si~$< 1$ could have profoundly different planetary geology with implications for habitability since vulcanism, silicate weathering, and the carbon cycle may all be altered by the differing mantle and crust properties \citep{2015ApJ...812...36F}.  

Planets of differing compositions will have slightly different mass-radius relations due to the varying equations of state of the rocks under pressure.  Estimates from models predict a change as small as 2\% in the radius and 6\% in the mass over the range of probable Mg/Si and Fe/Si ratios for an Earth-sized planet \citep{Zeng:2016cz}.  However, with better knowledge of these chemical ratios we can better constrain the core mass fractions and mantle radius fractions that influence mantle convection \citep{Unterborn:2016ig}, a key ingredient to planet habitability.  

\subsection{Intrinsic Distribution vs. Systematics}
The formal uncertainties on the abundances from \citet{Brewer:2016uh} account for the precision of fitting a model to observations and were already doubled to account for errors in their models.  However,  as uncertainties in the models are known to increase as temperatures and gravities diverge from solar values, we consider wether some of the observed spread in the C/O or Mg/Si ratios could be due to scatter from systematic errors.  For Sunlike stars, these model uncertainties should be minimized and only the statistical uncertainties should dominate. We evaluated the magnitude of these systematics by comparing the distributions of a restricted set of 344 stars with \logg~$> 4.0$, and \teff\ within $\pm 200$~K of the Sun.

\begin{figure*}[htpb!] 
   \centering
   \includegraphics[width=0.95\textwidth]{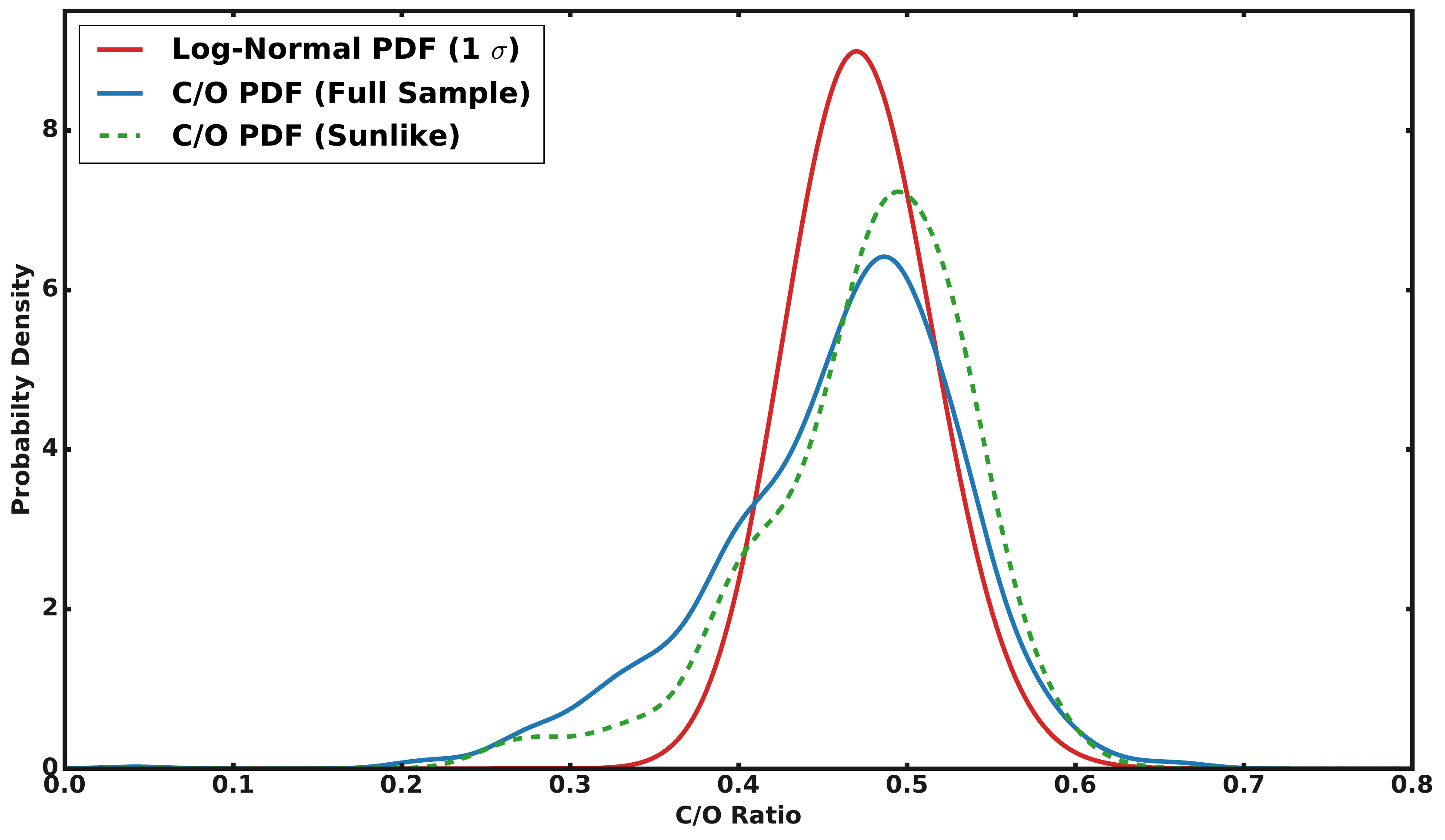} 
   \caption{The distribution of the full sample of C/O ratios (blue solid) compared to those of a sub-sample of 344 Sunlike stars (green dashed) and a log-normal distribution broadened by the formal uncertainties in [C/O] (red solid).  The distribution of Sunlike stars, where systematic uncertainties should be very small, is similar to that of the full distribution and more than twice as broad as the formal uncertainties. \\ }
   \label{fig:solar_comparison_CO}
\end{figure*}

Looking first at the C/O ratios (Figure \ref{fig:solar_comparison_CO}) we see that the distribution of Sunlike stars ifollows the same general distribution as the sample as a whole, but is slightly more centrally peaked.  This indicates that the systematic differences are small as we move away from solar values.  We also included the PDF of a log-normal distribution using the uncertainties in $[C/O]$ in Figure \ref{fig:solar_comparison_CO}.  The narrow distribution indicates that the formal uncertainties are also not responsible for the spread in the C/O ratios.  Multiplying the uncertainties by an additional factor of two would give a distribution similar to the higher C/O values but would still not explain the tail to lower values.

\begin{figure*}[htp!] 
   \centering
   \includegraphics[width=0.95\textwidth]{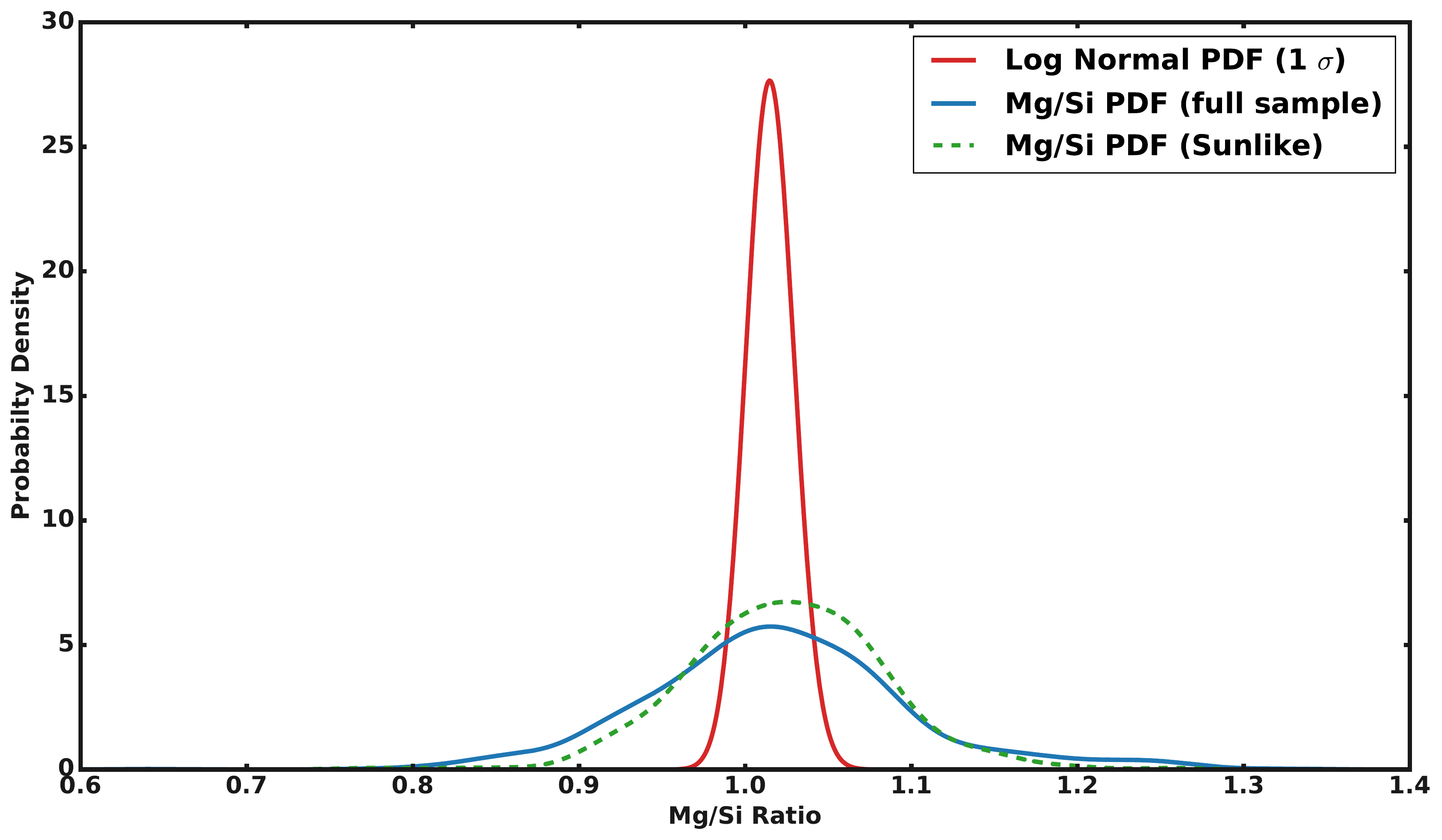} 
   \caption{As in Figure \ref{fig:solar_comparison_CO} but for Mg/Si. The full sample of Mg/Si ratios (blue solid) compared to those of a sub-sample of Sunlike stars (green dashed) and a log-normal distribution broadened by the formal uncertainties in [Mg/Si] (red solid).  The distribution of Sunlike stars is very similar to that of the full distribution and four times broader than the formal uncertainties.  The spread in Mg/Si is due largely to an intrinsic variation in the stellar abundances. \\ }
   \label{fig:solar_comparison_MgSi}
\end{figure*}

The uncertainties in [Mg/H] and [Si/H] are much smaller than for carbon and oxygen, though were derived in the same manner. In Figure \ref{fig:solar_comparison_MgSi} we have plotted the PDFs for the full sample, the sub-sample of Sunlike stars, and a log-normal PDF using the uncertainties in $[Mg/Si]$.  The PDF for the Sunlike stars is very similar to that of the sample as a whole, though slightly narrower. supporting our assertion that a distribution of Sunlike stars will exhibit smaller systematic errors.  However, the width of the Mg/Si distribution is too broad to be explained by modeling errors.  \citet{Brewer:2016uh} had already inflated the formal errors by a factor of two; we would need to scale this by an additional factor of four in order to reproduce a distribution that is similar to our restricted set of Sunlike stars.  To reproduce the width of the Mg/Si distribution for all of our stars, we would need to scale our errors up by a factor of five.  Without a physical motivation for such a large scaling factor for our formal errors, we conclude that the scatter in the distribution of Mg/Si has a cosmic origin that is intrinsic to our sample and not introduced by our analysis.

\subsection{Conclusions} \label{sec:conclusions}

We find that the carbon to oxygen ratio in the solar neighborhood is peaked just below the solar value at 0.47 with a steep drop off at super solar values.  There are no stars out of the 849 in our sample with C/O~$> 0.7$, implying that the formation of carbon-rich rocky planets is rare among stars represented by our sample.  The distribution of Mg/Si ratios for the stars peaks around 1.0 and the broad distribution suggests two populations of terrestrial planets having different rocky compositions: Earth-like and silicate rich.  Silicate rich planets very different than those in the solar system could form in 40\% of the systems in our sample. Though the differences in planet radius between these two compositions are currently unmeasurable, the stellar abundance ratios may be interesting for probing the range of rocky planet interiors or understanding the onset of plate tectonics and exoplanet geology.

%
\acknowledgments
The authors would like to thank Jessi Cisewski for help with the statistical analysis.  We would also like to thank the anonymous referee and statistics consultant who gave us valuable feedback for this paper. Data presented herein were obtained at the W. M. Keck Observatory from telescope time allocated to the National Aeronautics and Space Administration through the agency's scientific partnership with the California Institute of Technology and the University of California. The Observatory was made possible by the generous financial support of the W. M. Keck Foundation. This research has made use of the SIMBAD database, operated at CDS, Strasbourg, France 

The authors wish to recognize and acknowledge the very significant cultural role and reverence that the summit of Mauna Kea has always had within the indigenous Hawaiian community. We are most fortunate to have the opportunity to conduct observations from this mountain.

JMB and DAF thank NASA grant NNX15AF02G.

%
\bibliography{ms}

\typeout{get arXiv to do 4 passes: Label(s) may have changed. Rerun}
\end{document}